\documentclass[twoside,english,10pt]{JAC2001}

\usepackage{amssymb}
\usepackage{graphicx}
\usepackage{babel}
\usepackage{multicol}
\usepackage[centerlast]{caption2}

\begin{document}

\title{Effect of Cooling Water on Stability of NLC Linac Components}
\author{F.~Le~Pimpec, S.~Adiga (Stanford Univ.), F.~Asiri, G.~Bowden, D.~Dell'Orco,
E.~Doyle, \\ B.~McKee, A.~Seryi ~~SLAC, CA USA; H.~Carter,
C.~Boffo ~~FNAL, IL USA
\thanks{Work supported by the U.S. Department of Energy, Contract
  DE-AC03-76SF00515.}
 }
\maketitle

\begin{abstract}
Vertical vibration of linac components (accelerating structures,
girders and quadrupoles) in the NLC has been studied
experimentally and analytically. Effects such as structural
resonances and vibration caused by cooling water both in
accelerating structures and quadrupoles have been considered.
Experimental data has been compared with analytical predictions
and simulations using ANSYS. A design, incorporating the proper
decoupling of structure vibrations from the linac quadrupoles, is
being pursued.

\end{abstract}


\section{Introduction}
As the limit for the next generation of collider is again pushed
further, alignement of components and their stability are
critical. As part of the R\&D effort for the Next Linear Collider
(NLC), a program has developed to study the vibrations induced by
cooling water on the NLC linac components \cite{lepimpec:Epac02}
\cite{lepimpec:Linac02}. Similar studies are also underway at CERN
for the Compact Linear Collider (CLIC) \cite{clic531}.

An adequate flow of cooling water to the accelerating structures
is required in order to maintain the structure at the designated
operating temperature. This flow may cause vibration of the
structure and its supporting girder. The acceptable tolerance for
vibration of the structure itself is rather loose $\sim 1 \mu m$.
However our concern is that this vibration can couple to the linac
quadrupoles, where the vibration tolerance is 10~nm, either via
the beam pipe with its bellows or via the supports.

In this paper we will review the results obtained for the NLC
RF~structure and girder, those obtained for an electro magnetic
(EM) quadrupole when fed with water and finally the results of
coupling between the structure and the quadrupole when only the
structure is fed with water.

\section{Experimental setup}
\label{RFstructure}

\begin{figure}[tbph]
\begin{center}
\vspace{-0.3cm}
\includegraphics[width=0.55\textwidth,clip=]{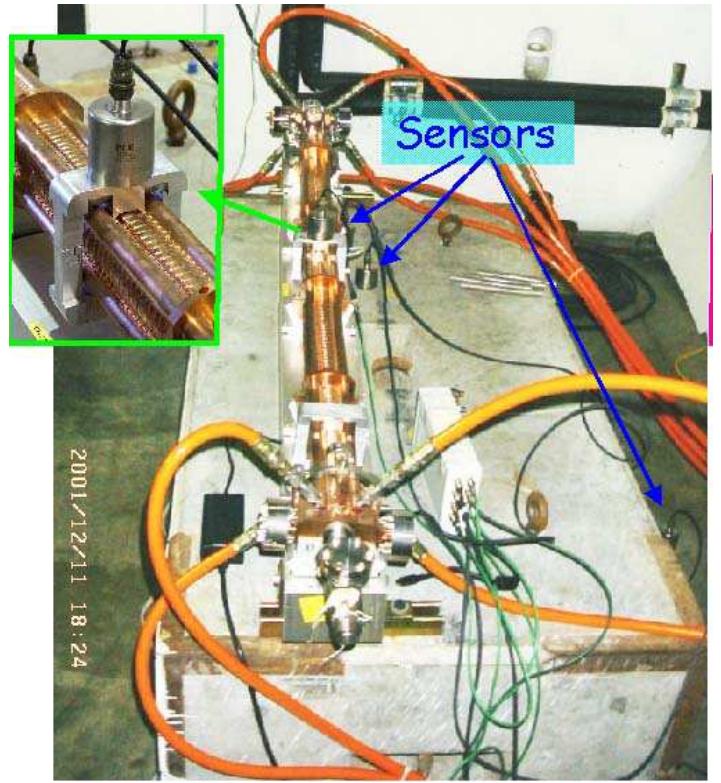}
\end{center}
\vspace{-0.3cm}
\caption{RF structure and girder vibration setup
in the NLCTA area. }
\vspace{-0.3cm}
\label{NLCTAvibsetup}
\end{figure}

\begin{figure}[tbph]
\begin{center}
\vspace{-0.3cm}
\includegraphics[width=0.50\textwidth,clip=]{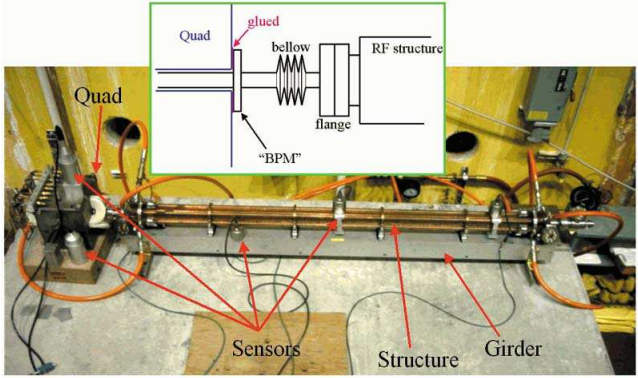}
\end{center}
\vspace{-0.3cm}
\caption{Experimental setup in
the SLD pit: RF structure-Girder system connected to the EM quad.}
\vspace{-0.3cm}
\label{QuadDDSsetup}
\end{figure}

The structure studied, DDS standing for Damped and Detuned
structure, is 1.8~m long, weight $\sim$100~kg and is supported by
a ``strongback'' (hollow aluminum box beam 4x6 inches) of the same
length, Fig.\ref{NLCTAvibsetup} and \ref{QuadDDSsetup}. In the
design, it was assumed that 3 such structures would be mounted on
a single 6~m long girder \cite{zdr}. It should be noted that the
NLC currently plans to use shorter RF structures than the one
studied \cite{adolphsen:Epac02}, probably 90 cm . The required
water flow (at 70MV/m loaded gradient) is about $\sim$1~$\ell$/s
for each structure (in total, through four cooling copper tubes).
The rise in temperature in copper, is given approximatively by :
$$\Delta T(^{\circ} C) \ = \
\frac{0.24*P(kW)}{Flow(\ell/s)}$$ The power loss in the structure
has been computed to be 4.5 kW/m and the required flow to obtain a
$\Delta T$, between RF on \& RF off for the structure, of 1
$^{\circ}$C is $\sim 1\ell/s$. For a filling time of 100~ns, a
raise of 0.5 $^{\circ}$C is equivalent to a 4$^{\circ}$ phase
shift at the structure output. This phase shift will induce less
than a half \% of energy variation for the beam ($\Delta
E/E<0.5$\%) if not compensate, and is acceptable. This 0.5
$^{\circ}$C range is the temperature tolerance for the entire
LINAC water cooling system \cite{adolphsen:TRC}.

In the first set of experiments, we measured the vibration induced
by different flow rates passing through the structure-girder
system, as shown in Fig.~\ref{NLCTAvibsetup}. The water was the
same as supplied to the NLC Test Accelerator (NLCTA) area. The RF
structure was mounted to a concrete block of $\sim$2225~kg. The
block was installed on rubber balls ($\sim$14Hz resonance) to
isolate it from the noisy floor of NLCTA. Vibration was monitored
by four piezo-accelerometers (PCB Piezotronics, INC model \#
393B31) and one piezo-transducer (PCB model \# 112A21) was used to
measure water pressure fluctuations. The diameter of each of the
four cooling pipes were 1.9~cm (OD). The flow of water and the
pressure were measured by a Venturi tube and by two manometers
installed at the input and output supply.

The second set of experiments were designed to study the vibration
caused only by internal turbulence. The structure-girder was in a
quieter place on the floor of the SLD (SLAC Large Detector)
collider hall and the cooling water was gravity-fed from a tank
located $\sim$18~m above. The structure-girder was bolted to a
$\sim$26T concrete block initially placed on a rubber mat and then
on sand (in the first configuration the block had resonance at
$\sim$35Hz which was decreased in the second case). The maximum
water flow through the structure was limited to about
1.1~$\ell$/s. The structure-girder was connected to the quad with
a bellows, and a simple mock-up of a BPM was connected (glued) to
the quadrupole Fig.\ref{QuadDDSsetup}. A primary vacuum,
sufficient for our tests, of $\sim 10^{-1}$ Torr could be obtain
in the vacuum pipe.

In each experiment, it was possible to feed the water to the
structure either by using the 4 tubes in parallel or by feeding
only 2 tubes at one end which were then connected in series to the
adjacent tubes at the other end. In the latter case, the flow in 2
tubes was in countered flow.

\section{Vibration of RF~Structure}

\begin{figure}[tbph]
\begin{center}
\vspace{-0.4cm}
\includegraphics[clip=,width=7.5cm,totalheight=5.5cm]{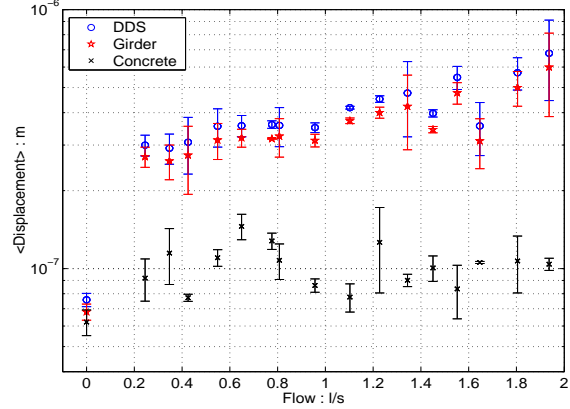}
\end{center}
\vspace{-0.4cm}
\caption{Average integrated displacement above 4Hz
of the RF structure (DDS), girder, and of the support (concrete
block) with NLCTA water supply.}
\label{NLCTAflowDDS}
\end{figure}

\begin{figure}[tbph]
\begin{center}
\vspace{-0.4cm}
\includegraphics[clip=,width=7.5cm,totalheight=5.5cm]{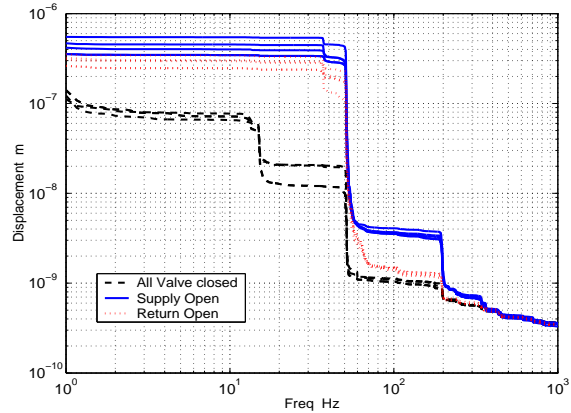}
\end{center}
\vspace{-0.4cm}
\caption{Integrated displacement spectrum of the
structure with only the supply open (blue solid line), only the
return open (red dotted line), and with all valves closed (black
dashed line). In all cases, there is no water flow.}
\vspace{-0.3cm}
\label{NLCTAvalveoff}
\end{figure}

Fig.\ref{NLCTAflowDDS} displays the results obtained in measuring
the vertical vibration induced by different flow rates passing
through the structure-girder system shown in
Fig.\ref{NLCTAvibsetup}. Note that the system considered is above
the turbulence threshold (Re$>$2000) when the flow
$>$~0.1~$\ell$/s. In Fig.\ref{NLCTAflowDDS} the water was supplied
by the NLC Test Accelerator (NLCTA) water system. In this case,
the displacement of the structure-girder is weakly dependant of
the flow in the structure because vibration is dominated by
hydraulic noise from water supply pumps. The supplying cooling
water has significant fluctuations of pressure in it (external
turbulence). Fig.\ref{NLCTAvalveoff} shows that vibration is not
dominated by water flow. Displacement is largest when cooling
inlet or outlet is open to supply pump but outlet or inlet is
closed (no flow). Merely opening either the supply or the return
valve, with the other valve closed, produced an integrated
displacement of 0.4~$\mu$m, equal to the maximum displacement
observed at nominal flow $\sim$1~$\ell$/s.

\begin{figure}[h]
\begin{center}
\vspace{-0.3cm}
\includegraphics[clip=,width=7.5cm,totalheight=5.5cm]{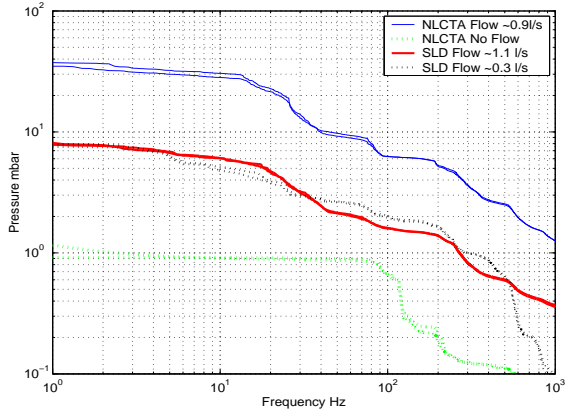}
\vspace{-0.5cm}
\end{center}
\caption{Integrated spectrum of incoming pressure fluctuations in
NLCTA water (top curves), in gravity fed water (middle curves),
and the sensor internal noise -- NLCTA case with all valves closed
(bottom curves).}
\vspace{-0.3cm}
\label{Presstransducer}
\end{figure}

Quantitatively, the incoming water can be characterized by the
spectrum of its pressure fluctuations. The integrated spectrum was
measured by a pressure piezo-transducer in the NLCTA setup as
shown in Fig.\ref{Presstransducer}. The incoming pressure causes
vibration of the RF structure through the force it exerts on the
surface of the cooling pipes (usually equal to cross-section of
the pipe). It is interesting to note that the spectrum of the
incoming NLCTA water is rather smooth and does not contain sharp
peaks typically associated with the rotational frequencies of
pumps. This indicates that the turbulence itself, and not the
pumps, was the cause of the pressure fluctuations.

The NLC cooling system will be designed so that pressure
fluctuations in the cooling water will be limited (if necessary,
by use of passive devices as typically done in industry
\cite{pulseguard}), internal turbulence will then dominate. Thus,
aiming to understand the contribution to vibration from internal
turbulence occurring inside the structure itself, we conducted the
second set of experiments.

\section{Vibration of RF structure and Coupling to Quadrupole }

Using the setup of Fig.\ref{QuadDDSsetup} we have studied the
vibration of RF structure versus flow, and the coupling of
vibration from the RF structure to the EM quadrupole in the case
when RF structure is cooled with gravity-fed water.

\begin{figure}[tbph]
\begin{center}
\vspace{-0.2cm}
\includegraphics[clip=,totalheight=5.5cm]{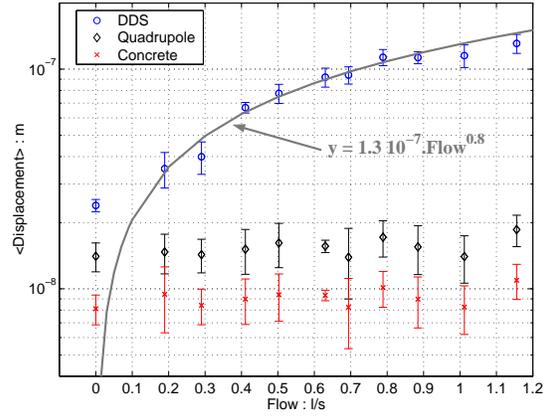}
\end{center}
\vspace{-0.4cm}
\caption{Average integrated displacement above
4Hz, with vacuum and gravity fed water. Fit by a power law (gray curve)}
\vspace{-0.3cm}
\label{SLDflowVac}
\end{figure}

\begin{figure}[tbph]
\begin{center}
\vspace{-0.2cm}
\includegraphics[clip=,totalheight=5.5cm]{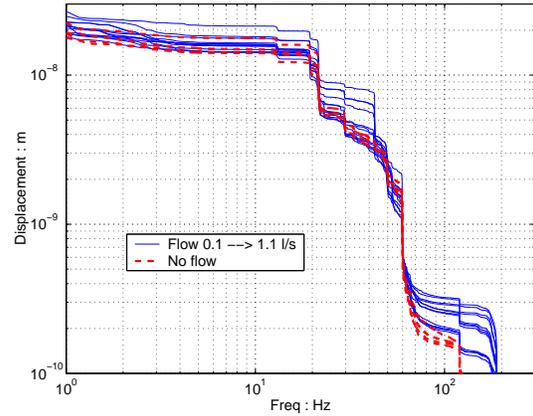}
\end{center}
\vspace{-0.4cm}
\caption{Quadrupole integrated displacement with
four different flows in the RF-structure, -SLD measurement.}
\vspace{-0.2cm}
\label{quadSLDspectrum}
\end{figure}

\begin{figure}[tbph]
\begin{center}
\includegraphics[clip=,totalheight=5.5cm]{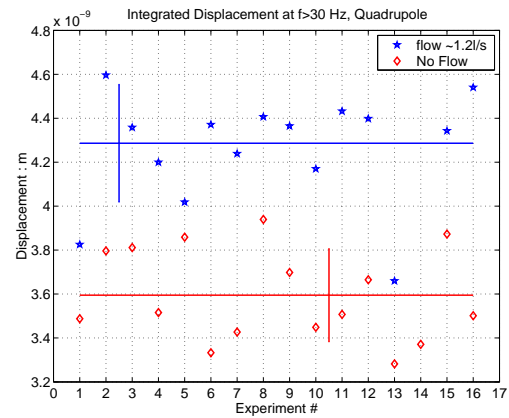}
\end{center}
\vspace{-0.4cm}
\caption{Coupling from the water cooled
RF-structure to the Quadrupole above 30Hz.}
\vspace{-0.4cm}
\label{quadcoupling30hz}
\end{figure}

Vibration of the RF structure versus flow is shown in
Fig.\ref{SLDflowVac}. In this case,  vibrations are caused mostly
by the internal turbulence occurring in the RF structure. At
nominal flow 1 $\ell/s$, vibration of the structure is
$\sim$110nm, compared to 350nm obtained with NLCTA cooling water
Fig.\ref{NLCTAflowDDS}. Additional vibrations of the quadrupole
are small. Performing multiple measurements with and without flow,
and analyzing spectra of quadrupole vibration
(Fig.\ref{quadSLDspectrum}), we found that the additional
vibration of the quadrupole due to cooling of RF structure above
30Hz is 2.4nm (obtained as $\sqrt{(4.3^2-3.6^2)}$, assuming
vibrations are uncorrelated), see Fig.\ref{quadcoupling30hz}.
Taking a lower cut frequency would be statistically uncertain, due
to high background noise of the concrete block at lower frequency.
These results suggest that coupling from RF structure to the
quadrupole is about 2\% in the current configuration. We also
investigated influence of vacuum in the RF structure (and possible
stiffening of the bellow) on this coupling. No noticeable
difference was observed with or without vacuum (the results
displayed in Fig.\ref{SLDflowVac} are obtained with a primary
vacuum of about ~10$^{-1}$~Torr in the structure-quadrupole
system). However, we have not yet studied how much coupling is due
to the bellows connection and how much due to transmission via
support and concrete.

One should also note that the present set up is simplified. In
particular, the quadrupole was placed on a small granite stand
(with shims to adjust the height), which was placed on concrete
block (without rigid connections). Such system had amplification
-- the quadrupole vibration is higher than the concrete as seen in
Fig.\ref{SLDflowVac}. This can be avoided in the real system.

\section{Simulation and theory}

\begin{figure}[tbph]
\begin{center}
\vspace{-0.3cm}
\includegraphics[clip=,width=6cm]{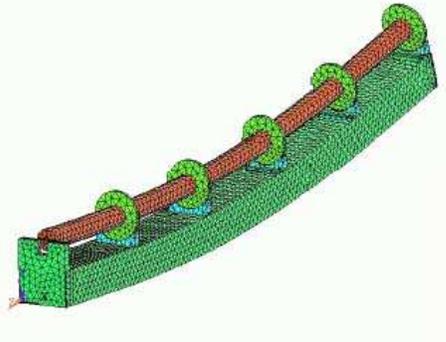}
\end{center}
\vspace{-0.4cm}
\caption{ANSYS simulation of the RF structure and
Aluminium girder, showing the lowest resonance mode.}
\vspace{-0.4cm}
\label{ANSYS49hz}
\end{figure}

We have shown in \cite{lepimpec:Epac02} that the vibration
spectrum of the girder-structure system exhibits a vertical
resonance at $\sim$52Hz Fig.\ref{NLCTAvalveoff}. The natural first
resonant frequency for such design, obtain through simulation with
ANSYS, is about $\sim$49~Hz. This calculated result is in good
agreement with measurements, and corresponds to simplest vertical
bending mode Fig.\ref{ANSYS49hz}. These simulations also indicate
that the second and the third modes are the horizontal dipole at
$\sim$69~Hz and vertical two-nodes mode at $\sim$117~Hz, while the
fourth resonance is torsional $\sim$146~Hz.

The driving forces (ground motion, water pressure and flow, …)
decrease rapidly with frequency. One possibility to further reduce
the vibration of the structure-girder system is to design a girder
which has a higher first resonant frequency. For further studies,
we have set a goal of increasing the lowest resonance frequency to
130~Hz and performed simulations to understand what modifications
this would require. One way to stiffen the girder is to increase
its dimensions. Simulations have shown that keeping the same
material and design but increasing the girder size (6"x4" to
10"x10") and the wall thickness (from 0.25" to 1") lead to a
lowest natural frequency of 120~Hz. Such big increase of the
resonance frequency may not be necessary, but the studies have
shown that significant improvement is possible with simple
modification of the girder design.

\begin{figure}[tbph]
\begin{center}
\vspace{-0.3cm}
\includegraphics[clip=,width=7.5cm,totalheight=5.5cm]{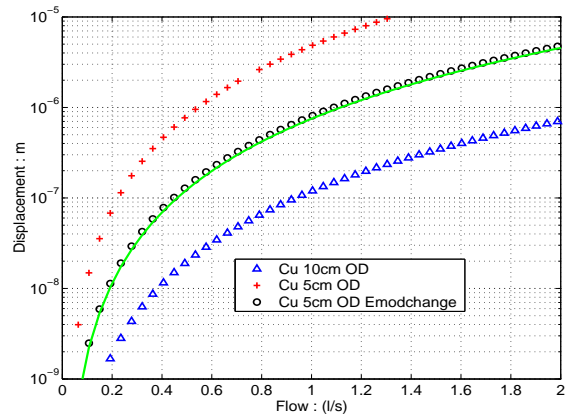}
\end{center}
\vspace{-0.4cm}
\caption{Theoretical calculated displacement of
the structure at different resonant frequencies. Triangles 47.97~Hz, crosses 24.78~Hz,
circles 49.55~Hz. The green line is a power fit.}
\vspace{-0.2cm}
\label{TheorieDDS}
\end{figure}

In parallel with the experiments and ANSYS analysis, a semi
analytical model is currently under development \cite{Sri:2002}.
Based on this model, an estimation of the vibration caused by the
flowing water has been computed for several cases. Results are
displayed in Fig.\ref{TheorieDDS}.

The model assumes a single pipe rigidly supported at each end,
with inner diameter (ID) equal to the ID of a cooling pipe, but
with a different thickness (OD). The actual OD of the structure is
$\sim$5cm. In order to scale to a system with 4 pipes, rather than
multiplying the result by 4, we assume that turbulence in
different pipes is independent. The total amplitude can be
estimated by multiplying the results by the square root of the
number of pipes. This vibrational analysis model is based on a
combination of observed and analytical techniques. The forcing
function \cite{Sri:2002} is measured in experimental model tests
and using this as input, the rms (root mean square) responses are
estimated based on probabilistic methods. In order to estimate
quantitatively the turbulence-induced vibrations, some empirical
data have been used. The mechanical damping factor of the
structure is taken equal to 0.01. The amplitude of vibrations is
inversely proportional to the square root of the damping factor.
The Reynolds number and boundary layer thickness relations are
based on empirical data. Depending on whether the flow has a
cavitating source or not, the appropriate empirical equations for
turbulent power spectral density need to be employed as well.

Since our model (single pipe) is simpler than the real
structure-girder system, we adjust the external diameter of the
modelling pipe to match at least the main mechanical
characteristics to those of the real system. In particular, we
match the first resonance frequency. This can be done by either
changing the OD or the Young modulus of the modelling pipe. The 3
resonant frequency for the curves in Fig.\ref{TheorieDDS} are
respectively for the cross 47.97~Hz, for the circles 24.78~Hz and
for the triangles 49.55~Hz. By artificially multiplying the Young
modulus by a factor 4, we are able, for a 5cm OD structure, to
obtain the right resonant frequency. We suppose that the best
representation, by our semi analytical model, of the system copper
structure - aluminium girder is given by taking the right
dimension of the structure (5~cm) and adjusting the elastic
modulus.

The theoretical results should be compared with the experimental
data obtained when the structure is fed with "quiet" water
Fig.\ref{SLDflowVac}. The comparison between the circles, and the
experimental results for the structure (labelled DDS), shows that
the semi-analytical model overestimates the vibration induced by
the flow, by a factor 8 at nominal flow 1~$\ell/s$. There is
agreement for flows between 0.2~$\ell/s$ and 0.4~$\ell/s$. It is
interesting to see that those curves can be fitted by a simple
power law :
\newline Displacement = $C_1$ * Flow $^{C_2}$;
\newline Applying this fit on our model, Fig.\ref{TheorieDDS} circles,
with the chosen $C_1$ $C_2$parameters gives :
\newline Displacement =  $7.5 \ 10^{-7}$ * Flow$^{2.6}$

\begin{figure}[tbph]
\begin{center}
\vspace{-0.3cm}
\includegraphics[clip=,width=7.5cm,totalheight=5.5cm]{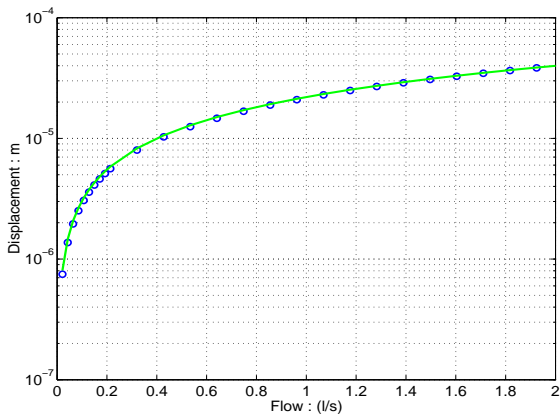}
\end{center}
\vspace{-0.4cm}
\caption{Theoretical Calculation from \protect{\cite{Schnell:2001kf}}
largely overestimate the vibration. The green line is a power fit.}
\vspace{-0.2cm}
\label{Theorieschnell}
\end{figure}

We also did compare our theoretical results with the analytical
formula suggested in \cite{Schnell:2001kf}. In this case we use
the formula assuming that the frequency is equal to the the
resonance f of the mechanical system. The frequency f is not
dependent of the flow as in \cite{Schnell:2001kf}.
Fig.\ref{Theorieschnell} is the results at a 50~Hz resonant
frequency. The simple theory derived from \cite{Schnell:2001kf}
gives a total displacement which is $\sim$200 times higher than
the reality at nominal flow ($10^{-7}$~m at 1~$\ell/s)$,
Fig.\ref{SLDflowVac}. Again the theory can be fitted with a power
law (green curve) :
\newline Displacement = $2.2\ 10^{-5}$ *
Flow$^{0.86}$
\newline In both cases, the power law fits theoretical estimation
which are derived from more complex formulae. However, the meaning
of the parameters $C_1$ and ${C_2}$ are not understood. We applied
this last power fit to our structure data (DDS),
Fig.\ref{SLDflowVac} gray curve, and by barely changing $C_2$
parameters and reducing $C_1$ by a factor 150, we were able to fit
them. It seems that the dependance of the vibration versus the
flow is a function with a square root dependency (0.8) rather than
a power dependency (2.5). In short, one theory predicts better the
amplitude, while the other better predicts the dependence on the
flow, and neither can be fully trusted.

\section{Vibration of EM Quadrupole}
\label{Quadresults}

\begin{figure}[htbp]
\begin{center}
\vspace{-0.2cm}
\includegraphics[clip=,width=0.45\textwidth]{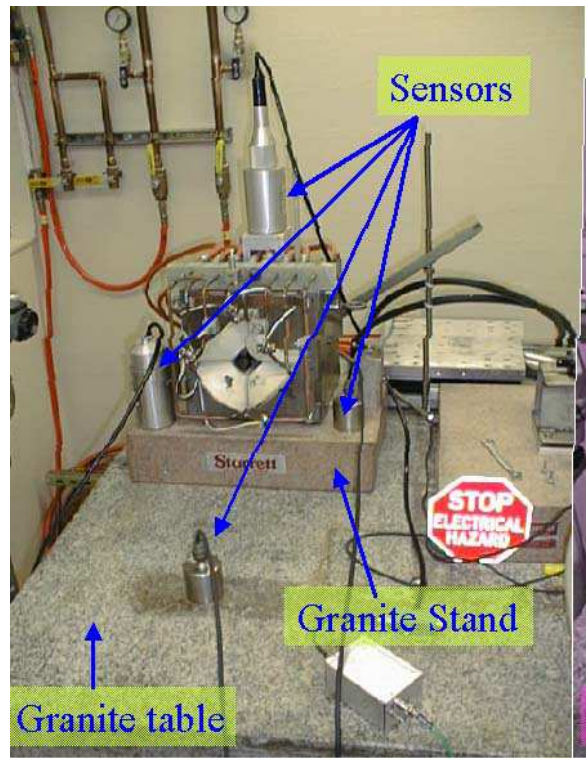}
\end{center}
\vspace{-0.4cm}
\caption{EM quadrupole vibration measurement setup. }
\vspace{-0.2cm}
\label{Quadsetup}
\end{figure}

The NLC project calls for maximal use of permanent magnet (PM)
quadrupoles which will not need cooling water. The electromagnet
quadrupoles (EM) however are also prototyped for NLC and we
studied vibration caused by cooling water in such an EM
quadrupole. The EM quadrupole was fed by a standard water supply
at a nominal flow of $\sim$0.1~$\ell/s$ obtained with pressure
difference of 8.5 bar. The quadrupole was installed on a granite
table Fig.\ref{Quadsetup}. The table was installed on rubber pads
to isolate the table from the high frequency vibration in the
noisy environment where measurements were performed. This reduced
the high frequency background, but significantly amplified
frequencies below 6-9~Hz, making it possible to study the effect
of cooling water on quadrupole vibration only above about 10Hz.

For f$>$20Hz, the vibration induced by the flow of
$\sim$0.1~$\ell/s$ in the quad is roughly 3.5~nm$\pm $0.25~nm
while 1~nm$\pm $0.25~nm at rest (averaged on several
measurements). Assuming that the additional vibration is
uncorrelated, the effect due to cooling water itself is:
$\sqrt{(3.52^2 - 1)}$ = 3.35~nm. The result is similar if a lower
cut off frequency (e.g. 15Hz) was considered, until below 10Hz
where statistical error becomes too big. Note that earlier studies
of FFTB quadrupole stability \cite{fftb_quad_vibro} have shown
that the effect of the cooling water is on a nanometer level as
well, for quadrupoles that were (in contrast to our study) also
properly placed on movers.

\section{Discussion}

With these data, we can estimate that in the pessimistic case, if
the cooling water will be similar to NLCTA (with similar pressure
fluctuations), vibration of the quadrupoles will scale to about
7.6nm (2.4nm*(350nm/110nm)) due to coupling to the RF structures.
In the case of EM quadrupoles, there will be about 3.3nm
additional due to cooling of the quadrupoles themselves, which in
total amounts to $\sqrt{(7.6^2 +3.3^2)}$ = 8.3nm. This value is
below the tolerance but has little margin. However, simple design
optimizations of girder or use of passive devices, such as damping
materials placed in or on the girder as well as in between the
water supply system and the structures in order to reduce the
fluctuations of pressure $\Delta$P seen in the supply system, are
expected to reduce these numbers considerably.

\begin{figure}[tbph]
\begin{center}
\vspace{-0.2cm}
\includegraphics[clip=,width=7.5cm,totalheight=5.5cm]{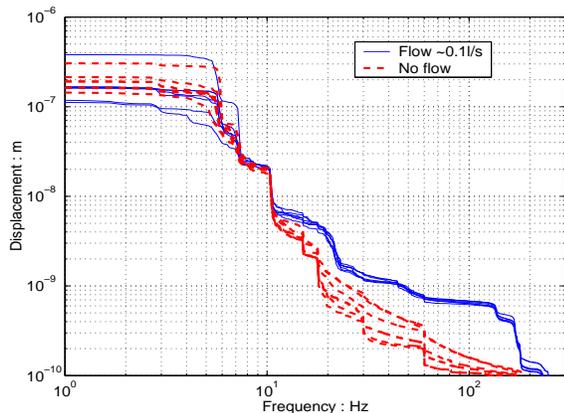}
\end{center}
\vspace{-0.4cm}
\caption{Integrated Displacement Spectrum of the
water induced vibration in the EM quadrupole at nominal flow. The
region below 10Hz is disturbed by resonances of the concrete table
installed on rubber pads. }
\vspace{-0.4cm}
\label{Quadresltspectrum}
\end{figure}

Among further studies of RF structure and quadrupole vibration
planned at FNAL and SLAC are: performing measurements in a quieter
place, quantifying lower frequency range; studying the case of
quadrupole placed on movers and realistic independent supports;
and continuing optimization of the system as a whole.

\section{Conclusion}

Cooling water can cause vibration of an accelerating structure
both through internal turbulence in the cooling pipes on the
structure, and through pressure fluctuations in the supply water
(external turbulence) \cite{lepimpec:Epac02}. The latter does not
depend on the flow rate through the structure and can be the
dominant source of vibration in practical situations.
\newline For the case studied, mechanical resonances of the structure-girder
assembly explain the measured amplitudes. Optimization of design
to increase resonance frequencies is expected to reduce vibration.
Coupling from RF structure to linac quadrupoles can occur via
bellows and the support, and was measured to be at the percent
level. Present studies suggest that the vibration tolerances for
the NLC linac quadrupoles are met, but without much margin.
Optimization of the girder design to improve its vibration
properties is highly desirable and will be pursued.

\section{Acknowledgments}
We would like to thank C. Adolphsen, R. Assmann, M. Breidenbach,
T. Raubenheimer, S. Redaelli, J. Sevilla, N. Solyak and C. Spencer
for help and useful discussions.


\end{document}